\documentclass[aps,twocolumn,showpacs,
superscriptaddress,groupedaddress,
floatfix,
twocolumn,aps,prd,amsmath,amssymb]{revtex4}

\usepackage[english]{babel}
\usepackage[utf8]{inputenc}
\usepackage{epsfig}
\usepackage{multirow}
\usepackage{graphicx}
\usepackage{graphics}
\usepackage{subfigure}
\usepackage{epstopdf}
\usepackage{float}
\usepackage{bm}
\usepackage{bbm}
\usepackage{braket}
\usepackage{lineno}
\usepackage{slashed}
\usepackage{hyperref}
\hypersetup{colorlinks=true,    linkcolor=red,filecolor=black,
citecolor=blue}

\begin{document}

\title{Influence of the four-fermion interactions in (2+1)D massive electrons system}

\author{Luis Fernández}
\email{luis.aguilar@icen.ufpa.br}
\affiliation{Faculdade de F\'isica, Universidade Federal do Par\'a, 66075-110 Bel\'em, Par\'a, Brazil}

\author{Van Sérgio Alves}
\email{vansergi@ufpa.br}
\affiliation{Faculdade de F\'isica, Universidade Federal do Par\'a, 66075-110 Bel\'em, Par\'a, Brazil}

\author{M. Gomes}
\email{mgomes@fma.if.usp.br}
\affiliation{Instituto de F\'isica, Universidade de S\~ao Paulo, Caixa Postal 66318, 05315-970 S\~ao Paulo, S\~ao Paulo, Brazil}

\author{Leandro O. Nascimento}
\email{lon@ufpa.br}
\affiliation{Faculdade de Ci\^encias Naturais, Universidade Federal do Par\'a, C.P. 69900-000 Breves, Par\'a, Brazil}

\author{Francisco Peña}
\email{francisco.pena@ufrontera.cl}
\affiliation{Departamento de Ciencias F\'isicas, Facultad de Ingenier\'ia y Ciencias, Univesidad de La Frontera, Avenida Francisco Salazar 01145, Casilla 54-D, Temuco, Chile}

\date{\today}

\begin{abstract}
{The description of the electromagnetic interaction in two-dimensional Dirac materials, such as graphene and transition-metal dichalcogenides, in which electrons move in the plane and interact via virtual photons in 3d, leads naturally  to the emergence of a projected non-local theory, called pseudo-quantum electrodynamics (PQED), as an effective model suitable for describing electromagnetic interaction in these systems. In this work, we investigate the role of a complete set of four-fermion interactions in the renormalization group functions when we coupled it with the anisotropic version of  massive PQED, where we take into account  the fact that the Fermi velocity is not equal to the light velocity. We calculate the electron self-energy in the dominant order in the $1/N$ expansion in the regime where $m ^ 2 \ll p ^ 2$. We show that the Fermi velocity renormalization is insensitive to the presence of quartic fermionic interactions, whereas the renormalized mass may have two different asymptotic behaviors at the high-density limit, which means a high-energy scale.}
\end{abstract}

\maketitle

\section{INTRODUCTION}

Four-fermion interactions have been extensively studied in the literature, both for understanding conceptual aspects of quantum field theory as well as for applications in condensed matter physics. In particular, the Thirring \cite{Thirring} and Nambu-Jona-Lasinio \cite{NJL} models show a rich connection between the phenomenon of superconductivity and elementary particle physics. The latter has also been used for studying quantum chromodynamics at the low-energy limit \cite{Cahil,Hatsuda}. Although four-fermion interactions are perturbatively non-renormalizable in a space-time $D>2$, in the sense of general power counting rules \cite{Dyson}, they become renormalizable when we use the $1/N$ expansion in $D = 3$ \cite{four1/N}. Indeed, the incorporation of vacuum polarization effects provide a better behavior for the Green functions in the ultraviolet regime. Therefore, both the Gross-Neveu \cite{GN} and Thirring \cite{MGomes} interactions may be renormalizable in $D=3$. Usually, in order to perform the $1/N$ expansion, it is used a Hubbard-Stratonovich transformation \cite{hubbard} through the introduction of an auxiliary field, which has no dynamics at the tree level.

It is well known that the quasiparticle excitations in two-dimensional materials at the honeycomb lattice (such as graphene \cite{Grafeno}, silicene \cite{Siliceno}, and transition metal dichalcogenides \cite{TMD}) behave as Dirac-like fermions (either massless or massive). Hence, the four-fermion interactions also become relevant, as an attempt to obtain a more complete description of these systems, within a quantum-field-theory approach. Indeed, this more realistic description should take into account some of the microscopic interactions that, such as disorder/impurity, may emerge in these materials. Because the auxiliary fields obey the same properties as the random disorder/impurities interactions, as discussed in Refs.\cite{Liu,Wang}, hence, we can relate these properties of the materials with the four-fermion interactions, within the low-energy limit. Furthermore, it is also very useful to consider the electromagnetic interactions in the plane, which may be effectively described by the pseudo-quantum electrodynamics model \cite{marino}.  

In a previous work we analyzed the effect of the electromagnetic interaction on the  renormalization of the mass  gap of electrons moving in a plane subject also to impurities simulated by a Gross-Neveu like self-interaction \cite{fernandez}. Without the four-fermion interaction, we derived results that are in excellent agreement with experimental measurements of the band gap for WSe$_2$ \cite{WSe} and MoS$_2$ \cite{MoS}. We found also that, although the presence of the Gross-Neveu like interaction does not alter the renormalization of the Fermi velocity \cite{vozmediano},  it provides an ultraviolet fixed point in terms of an effective fine-structure constant, so that the renormalized mass has different behaviors below and above it. 

In this paper we extend the  investigation  presented in \cite{fernandez}   by considering 
the generalized four-fermion interactions with $O(4)$ symmetry.

The remainder of this paper is organized as follow. In
Sec. II we present our model, notation, and perform the expansion $1/N$ through the Hubbard-Stratonovich transformation, which allow us to define the Feynman rules.  In Sec. III  we calculate the propagators of the gauge and auxiliary fields in the dominant order in $1/N$ in the regime where $m ^2\ll p^2$.  In
Sec. IV we calculate the electron self-energy due to electromagnetic and the four-fermions interactions, taking into account the effect of the polarization tensor obtained in the previous section.  The derivation of the renormalization group functions and the effect of each four-fermion interaction on the renormalized mass are shown in Sec.V.  In Sec. VI we review our main results and conclusions. Some details about the derivation of the polarization tensor, due to the  four-fermion interactions, are given in Appendix \ref{A}.

\section{PSUEDO-QUANTUM ELECTRODYNAMICS WITH FOUR-FERMIONS INTERACTION}

We consider the PQED model \cite{marino} with a complete set of independent four-fermion interactions in (2+1)D \cite{gomes1}. The Euclidean action reads 
\begin{equation}
 \begin{split}
\mathcal{L}&=\frac{1}{2}\frac{F^{\mu \nu} F_{\mu \nu}}{\sqrt{-\Box}} +\bar{\psi}_a\left(\dot{\imath} \gamma^{\mu}D_{\mu}-m\right)\psi_a  - \xi \frac{\left(\partial_{\mu} A^{\mu}\right)^2}{\sqrt{-\Box}}  \\ & -\sum^8_{l=1} \frac{G_l}{2}\left(\bar{\psi}_a \Gamma^l \psi_a\right)^2,
\label{model}
\end{split}
 \end{equation} 
where $F_{\mu \nu} = \partial_{\mu} A_{\nu} - \partial_{\nu} A_{\mu}$ is the field intensity tensor of the gauge field $A_{\mu}$, $\Box$ is the d'Alembertian operator, $\psi_a$  is the Dirac field, and $a= 1,..., N$ is the flavor index. For electrons in the honeycomb lattice, we may use the representation for matter field as $\psi_a^{\dagger}=(\psi^{*}_{A \uparrow}, \psi^{*}_{A\downarrow}, \psi^{*}_{B\uparrow}, \psi^{*}_{B\downarrow})_a$, where $(A,B)$ and $(\uparrow,\downarrow)$ are the sublattices and spins, respectively. Therefore, one finds $a=K, K'$ and $N=2$ that describes the valley degeneracy. Here, we perform all of the calculations for an arbitrary value of $N$ \cite{Gfactor, libroMarino}. Furthermore, $m$ is the Dirac mass, $e$ is the electric charge, $\xi$ is the gauge-fixing parameter, $G_l=\{G_1,...,G_8\}$ are the coupling constants of the four-fermion interactions where $l=1,...,8$ is an index describing each self-interaction, $\Gamma_l= \{ \mathbbm{1}, \gamma^\mu, \gamma^3, \gamma^5, \gamma^\mu \gamma^3, \gamma^\mu\gamma^5, \gamma^3\gamma^5, \gamma^\mu \gamma^3\gamma^5 \} $ are their corresponding matrices, $\gamma^\mu$ are the Dirac matrices in the $4 \times 4$ representation, whose algebra is given by $\left\lbrace \gamma^{\mu}, \gamma^{\nu} \right\rbrace = - 2 \delta^{\mu \nu}$, and $\gamma^{\mu} D_\mu$=$\gamma^0\partial_0 + v_F \gamma^i \partial_i +e \gamma^\mu A_{\mu}$ is the Dirac operator after we perform the minimal coupling with $A_\mu$. Our matrix representation follows the definition given in Ref.~\cite{wang}. Thus, our Dirac matrices are anti-hermitian: $(\gamma_0,\gamma_{1},\gamma_{2})=(i\sigma_{3},i\sigma_{1},i\sigma_{2})\otimes,\sigma_{3}), \gamma_{3}= I\otimes \sigma_1$ and $\gamma_{5}\equiv I\otimes \sigma_{2}$ so that $(\gamma_{\mu})^{2}=-1$, and $\gamma_{5}$ is Hermitian. Furthermore, we shall use the natural system of units, where $\hbar=c=1$. Because $[G_l]=-1$, the model in Eq.~(\ref{model}) is not renormalizable in the perturbative expansion, but it is in the large-$N$ expansion. Hence, we shall consider the large-$N$ expansion from now on.

The first step is to introduce the $N$ parameter into the action through a scaling of the coupling constants, given by $e \rightarrow e/\sqrt{N}$ and $G_l \rightarrow G_l/N$ for a fixed $e$ and $G_l$, respectively. Thereafter,  we use a Hubbard-Stratonovich transform in the four-fermion interactions, given by
\begin{equation}
\begin{split}
\frac{G_l}{2N}\left(\bar{\psi}_a \Gamma^l \psi_a\right)^2 &\rightarrow \frac{G_l}{2N}\left(\bar{\psi}_a \Gamma^l \psi_a\right)^2 \\
&- \frac{N}{2 G_{1}} \left[\varphi_1 - \frac{G_{1}}{N}\bar{\psi}_a\Gamma_1 \psi_a\right]^2-...  \\
&- \frac{N}{2 G_{8}} \left[\varphi_8 - \frac{G_{8}}{N}\bar{\psi}_a\Gamma_8 \psi_a\right]^2, \label{HS}
\end{split}
\end{equation}
where 
\begin{equation}
\varphi_l=\{\varphi_{\mathbbm{1}},\varphi_{\gamma^\mu},\varphi_{\gamma^3},\varphi_{\gamma^5},\varphi_{\gamma^\mu \gamma^3},\varphi_{\gamma^\mu\gamma^5},\varphi_{\gamma^3\gamma^5},\varphi_{\gamma^\mu \gamma^3\gamma^5}\}\nonumber
\end{equation}
 is a set of auxiliary fields for each kind of interaction. Note that, for the sake of simplicity, we applied the notation $\varphi_l=\{\varphi_1,...,\varphi_8\}$ in Eq.~(\ref{HS}). Using Eq.~(\ref{HS}) in Eq.~(\ref{model}), one finds the motion equation for the auxiliary fields, namely, $\varphi_l= G_{l}\bar{\psi}_a\Gamma_l \psi_a/N$ at classical level for each $l=1,...,8$ (there is no sum over $l$ in the rhs of this equation). Furthermore, we also obtain the action
\begin{equation}
 \begin{split}
\mathcal{L}&=\frac{1}{2}\frac{F^{\mu \nu} F_{\mu \nu}}{\sqrt{-\Box}} +\bar{\psi}\left(\dot{\imath} \gamma^{0}\partial_{0}+\dot{\imath} v_F \gamma^{i}\partial_{i}-m\right)\psi + \\ &+\frac{e}{\sqrt{N}}\bar{\psi}\gamma^{\mu}\psi A_{\mu} - \xi \frac{\left(\partial_{\mu} A^{\mu}\right)^2}{\sqrt{-\Box}}   \\ &  +\sum^8_{l=1} \left[\frac{N}{2 G_{l}} \varphi_l^2   - \varphi_l \bar{\psi}\Gamma_l\psi\right]. \label{model1}
\end{split}
\end{equation}
Next, we realize a simple shift in the auxiliary field, namely, $\varphi_l\rightarrow \sigma_{0,l}+\varphi_l/\sqrt{N}$ such that $\sigma_{0,l}=\braket{\varphi_l}$ is the vacuum expectation value of $\varphi_l$. Using this transform in Eq.~(\ref{model1}), we have
\begin{equation}
 \begin{split}
\mathcal{L}&=\frac{1}{2}\frac{F^{\mu \nu} F_{\mu \nu}}{\sqrt{-\Box}}\! +\!\bar{\psi}\left(\dot{\imath} \gamma^{0}\partial_{0}\!+\!\dot{\imath} v_F \gamma^{i}\partial_{i}-m\!-\!\sigma_{0,l}\Gamma_l\right)\!\psi  \\ &+\frac{e}{\sqrt{N}}\bar{\psi}\gamma^{\mu}\psi A_{\mu} - \xi \frac{\left(\partial_{\mu} A^{\mu}\right)^2}{\sqrt{-\Box}} +\sum^8_{l=1} \left[\frac{N}{2 G_{l}} \sigma_{0,l}^2 \right]  \\ &  + \sum^8_{l=1} \left[\frac{1}{2 G_{l}} \varphi_l^2   - \frac{\sqrt{N}}{G_{l}} \sigma_{0,l} \varphi_l-\frac{1}{\sqrt{N}}\varphi_l \bar{\psi}\Gamma_l\psi\right]. \label{model2}
\end{split}
\end{equation}
One advantage of Eq. \eqref{model2} is that for $m=0$ it clearly separates the analysis into  phases, i.e, one with no spontaneous symmetry breaking where $\sigma_{0,l}=0$ and other phases with some broken symmetry $\sigma_{0,l}\neq 0$. In particular, a phase with chiral symmetry breaking , i.e, $\sigma_{0,1} \neq 0$ has been discussed in Ref.~\cite{fernandez}. 
Next, let us define the Feynman rules. The gauge-field propagator in Eq. \eqref{model1} reads
\begin{equation}
\Delta^0_{\mu \nu}(p)= \frac{1}{2 \sqrt{p}}\left[\delta_{\mu \nu}-\left(1-\frac{1}{\xi}\right)\frac{p_\mu p_\nu}{p^2}\right],
\label{fotonbare}
\end{equation}
while the fermion propagator is given by
\begin{equation}
S_F(p)=-\frac{1}{\gamma^0p_0+v_F\gamma^ip_i -m},
\label{fermionbare}
\end{equation}
and,in the  tree approximation, the propagator for the  auxiliary-field $\varphi_{l}$ is
\begin{equation}
\left(\Delta^0_{{\varphi_l}}\right)=\left(\frac{1}{G_{l}}\right)^{-1}.
\end{equation}
The electromagnetic and trilinear vertices interactions are given by $e/\sqrt{N}$ and $1/\sqrt{N}$, respectively. Next, we shall calculate the quantum corrections, within the large-$N$ approximation, for the field propagators.

\section{FULL PROPAGATORS}
\subsection{Gauge-field propagator}
The full gauge-field propagator, in the dominant order of $1/N$, is written as \cite{fernandez} 
\begin{equation}
\Delta_{\mu \nu}(p)= \Delta_{\mu \nu}^0(p)+\Delta_{\mu \alpha}^0(p)\Pi^{\alpha \beta}(p)\Delta_{\beta \nu}^0(p)+ \cdots,
\label{fullfot}
\end{equation}
where $\Pi^{\mu\nu}(p)$ is the vacuum polarization tensor, namely,
\begin{equation}
\Pi^{\mu \nu}(p)= -\frac{e^2}{N}{ \rm Tr} \int \frac{d^3 k}{(2 \pi)^3} \gamma^{\mu}S_F(p+k)\gamma^\nu S_F(k).
\end{equation}
In the static limit, we only need the component $\Pi^{00}(p)$ given by
\begin{equation}
\Pi^{00}(\textbf{p}^2)=-\frac{e^2}{8} \frac{\textbf{p}^2}{\sqrt{p_0^2 + v_F^2 \textbf{p}^2}} \label{pi00}
\end{equation}
in the small-mass limit $m^2\ll p^2 $. Using Eq.~(\ref{pi00}) in Eq. \eqref{fullfot}, we find 
\begin{equation}
\Delta_{00}(\textbf{p}^2)= \left(2 \sqrt{\textbf{p}^2}+\frac{e^2}{8} \frac{\textbf{p}^2}{\sqrt{p_0^2 + v_F^2 \textbf{p}^2}}\right)^{-1}.
\label{fullfot2}
\end{equation}
This agrees with the result in Ref.~\cite{fernandez}. 

\subsection{Auxiliary-field propagators}
The quantum corrections for the auxiliary fields $\varphi_l$ may be obtained through the effective action $S_{\rm eff}$. This is accomplished from Eq.~(\ref{model2}) by  integrating out the matter field. After expanding $S_{\rm eff}$ for large-$N$, we find
\begin{equation}
S_{\rm eff}[\varphi_l] = \sqrt{N}S_1[\varphi_l] + S_2[\varphi_l] + \cdots, \label{SeffN}
\end{equation}
where
\begin{equation}
\begin{split}
S_1&= {\rm Tr} \left[\left( \dot{\imath}\gamma^0 \partial_0 + \dot{\imath} v_F \gamma^i \partial_i -m-\sigma_{o,l} \Gamma^l\right)^{-1} \right. \\ & \left. \times \left(\sum_l \varphi_l \Gamma^l \right)\right] 
 +  \sum_l \frac{1}{G_{l}}\sigma_{0,l} \varphi_l  
\label{S1}
\end{split}
\end{equation}
and
\begin{equation}
\begin{split}
S_2 &=\frac{1}{2}{\rm Tr}\left[\left\lbrace  \left( \dot{\imath}\gamma^0 \partial_0 + \dot{\imath} v_F \gamma^i \partial_i -m-\sigma_{o,l} \Gamma^l\right)^{-1} \right. \right. \\ & \left. \left. \times \left( \sum_l\varphi_l \Gamma^l \right) \right\rbrace^2 \right]  + \int d^3 x \sum_l \frac{1}{2G_l}\varphi_l^2.
\label{S2}
\end{split}
\end{equation}
Note that $S_1$ in Eq.~(\ref{S1}) may be written as $S_1=\sum_l \varphi_l S_l[\sigma_{0,l},G_l]$. On the other hand, we have that $S_1=0$ which implies a convergent effective action in Eq.~(\ref{SeffN}). This yields a set of gap equations $S_l[\sigma_{0,l},G_l]=0$ for each $l$, giving a nontrivial relation between the values of $\sigma_{0,l}$ and the coupling constants $G_l$. However, for $\sigma_{0,l}=0$ and $m=0$ these gap equations are automatically satisfied.

\begin{figure}[H]
\centering
\includegraphics[scale=0.9]{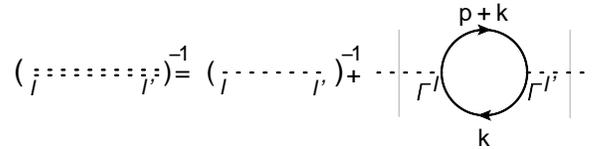}
\caption{The large-$N$ approximation for the full auxiliary-field propagators. The full propagators of the auxiliary fields are represented by the double-dashed line with the subscripts $(l,l')$, which are meant to describe their different tensorial structure. The continuum line is the propagator of the fermion field.}\label{fullaux}
\end{figure}

Next, it is convenient to write Eq. \eqref{S2} as 
\begin{equation}
 S_2= \frac{1}{2}\int d^3 x \, d^3y \,\, \varphi_l(x)\, \Gamma^{ll'}(x-y)\,\varphi_{l'}(y),
 \end{equation} 
providing the auxiliary-field propagator $\Gamma^{ll'}(x-y)^{-1}$. This, in the momentum space, is \textit{schematically} written as
\begin{equation}
\left[\Delta^{ll'}_{\{\varphi_l\}}(p)\right]^{-1}=\Gamma^{ll'}_{\{\varphi_l\}}(p) = \frac{1}{G_{\{\varphi_l\}}}\delta^{ll'} - \Pi^{ll'}_{\{\varphi_l\}}(p).
\label{fullauxiliary}
\end{equation}
At this point, we must be careful with our notation in order to avoid any misunderstanding. Indeed, the kind of indexes $(ll')$ we have in Eq.~(\ref{fullauxiliary}) depends on the kind of auxiliary field $\{\varphi_l\}$ we want to consider. For instance, $\varphi_1=\varphi_{\mathbbm{1}}\rightarrow \sigma$ is a scalar field, hence, $\delta^{ll'}$ only means an unity. Nevertheless, we may consider the second auxiliary field, which is actually $\varphi_2=\varphi_{\gamma_{\mu}}\rightarrow {\cal A}_\mu$  a vector field. In this case, we must consider that $\delta^{ll'}\rightarrow \delta^{\mu\nu}$, where we replace $(ll')$ by two Lorentz indexes, i.e, $(ll')\rightarrow (\mu\nu)$, such that we find a propagator $\Delta^{\mu\nu}_{\{\varphi_{\mu}\}}(p)$, as expected. The main rule is that for a generic auxiliary field $\{\varphi_l\}$, one must have a scalar quantity $S_2 \propto \varphi_l \Gamma^{ll'} \varphi_{l'}$, which, therefore, fixes the tensorial structure of $\Gamma^{ll'}$. We represent the full propagator of the auxiliary fields in Fig. \ref{fullaux}.

The different self-energies for each auxiliary field read 
\begin{equation}
\Pi^{ll'}_{\{\varphi_l\}}(p) = - {\rm Tr} \int \frac{d^3 k}{(2 \pi)^3} \Gamma^l S_F(p+k) \Gamma^{l'} S_F(k). \label{auxprop}
\end{equation}
We consider the $4 \times 4$ representation of the Dirac matrices, whose trace operations are detailed in appendix \ref{A}. Because of the Lorentz symmetry in the Dirac matrices, we perform a redefinition of the external momentum  as $v_F p_i \rightarrow \bar{p}_i$,  such that $\bar{p}_{\mu} =(p_0, \bar{p}_i)$. Furthermore, for the sake of consistency, we also change the spatial-variable of the loop integral as $v_F k_i \rightarrow k_i$, which implies that $d^3 k \rightarrow d^3 k / v_F^2$. Therefore,
\begin{equation}
\Pi^{ll'}_{\{\varphi_l\}}(\bar{p})= -\frac{1}{v_F^2} {\rm Tr} \int \frac{d^3 k}{(2 \pi)^3} \Gamma^l S_F(\bar{p}+k) \Gamma^{l'} S_F(k).
\label{Polarization4f}
\end{equation}
It is clear that the only difference, between the different four-fermion interactions, is the vertex structure $\Gamma_l$ (and $\Gamma_{l'}$) in Eq.~(\ref{Polarization4f}).

\section{The Electron Self-Energy}

We assume the symmetric phase, where $\sigma_{0,l}=0$. This phase is promptly obtained from Eq.~(\ref{model2}) by using $\sigma_{0,l}=0$. Using the Feynman parametrization and the dimensional regularization (See Appendix \ref{A}), we obtain the self-energies for all of the auxiliary fields $\varphi_l$ which for higher momenta, is given by
\begin{align}
\Gamma_{\varphi_\mathbbm{1}}(\bar{p}) & = \frac{1}{G_{\varphi_\mathbbm{1}}}+ \frac{\sqrt{\bar{p}^2}}{4 v_F^2}, \\
\Gamma_{\varphi_\mu}^{\mu \nu}(\bar{p}) & = \left(\frac{1}{G_{\varphi_\mu}}+\frac{\sqrt{\bar{p}^2}}{8 v_F^2}\right)\mathbb{\bar{P}}^{\mu \nu} + \frac{1}{G_{\varphi_\mu}}\frac{\bar{p}^\mu \bar{p}^\nu}{\bar{p}^2},\label{vector1}\\
\Gamma_{\varphi_{3(5)}}(\bar{p}) & = \frac{1}{G_{\varphi_{3(5)}}}- \frac{\sqrt{\bar{p}^2}}{4 v_F^2},\\
\Gamma^{\mu \nu}_{\varphi_{\mu3(5)}}(\bar{p}) & =\!\! \left(\!\frac{1}{G_{\varphi_{\mu3(5)}}}\!+\!\frac{\sqrt{\bar{p}^2}}{8 v_F^2}\right)\mathbb{\bar{P}}^{\mu \nu} + \frac{1}{G_{\varphi_{\mu3(5)}}}\frac{\bar{p}^\mu \bar{p}^\nu}{\bar{p}^2},\label{vector2}\\
\Gamma_{\varphi_{35}}(\bar{p}) & = \frac{1}{G_{\varphi_{35}}}+ \frac{\sqrt{\bar{p}^2}}{4 v_F^2},
\end{align}
and
\begin{equation}
\Gamma^{\mu \nu}_{\varphi_{\mu35}}(\bar{p})  =\!\! \left(\!\frac{1}{G_{\varphi_{\mu35}}}\!-\!\frac{\sqrt{\bar{p}^2}}{8 v_F^2}\right)\mathbb{\bar{P}}^{\mu \nu} + \frac{1}{G_{\varphi_{\mu35}}}\frac{\bar{p}^\mu \bar{p}^\nu}{\bar{p}^2}.\label{vector3}
\end{equation}
The subscription $3(5)$ means that the result holds for both $\varphi_3$ and $\varphi_5$ fields, for example. Furthermore, the standard projection tensor $\mathbb{\bar{P}}^{\mu \nu}$ reads
\begin{equation}
\mathbb{\bar{P}}^{\mu \nu}=\delta^{\mu\nu}-\frac{\bar{p}^\mu \bar{p}^\nu}{\bar{p}^2}.
\end{equation}

It should be noticed the bad  ultraviolet behavior of the longitudinal part of the two point proper function involving a vectorial field, namely the longitudinal parts in Eqs. (\ref{vector1}), (\ref{vector2}), and (\ref{vector3}). Of course, these bad behaviors are innocuous if the corresponding currents are conserved. In any case, this fact is only relevant for calculating the correction in order  $1/N^2$. If we consider only the transversal part of these propagators, the generalized model (\ref{model1}) is power counting renormalizable with divergences being eliminated by reparametrizations of the fields 
and of  the mass of the fermion field. In what follows we will discuss in detail the divergences in the fermion self-energy.
\subsection{The Fermion Self-enegy}

Having the gauge and auxiliary-field propagators, we may calculate the fermion self-energy. This also may be decomposed into two terms, one due to the gauge field and the other due to the auxiliary fields.

\begin{figure}[H]
\centering
\includegraphics[scale=0.9]{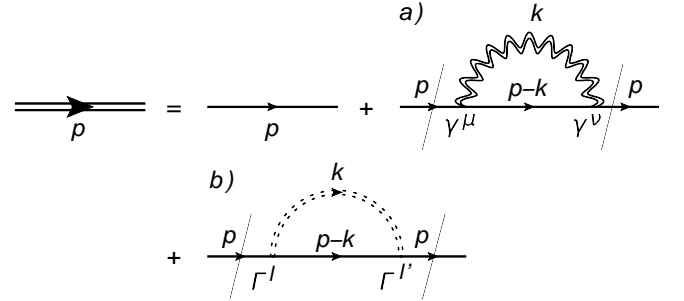}
\caption{The full fermion propagator up to the dominant order $1/N$. The full fermion propagator is represented by the double continuos lines. $a)$ The self-energy due to the interaction between the gauge and fermion fields. $b)$ The general structure of the self-energies due to each auxiliary-field propagator and the fermion field.}\label{selfenergy1}
\end{figure}

\subsubsection{Self-energy due to the gauge field}
The fermion self-energy due to the gauge field is shown in Fig \ref{selfenergy1}\textcolor{red}{.a} and its analytical expression is given by
\begin{equation}
\Sigma_{A_\mu}(p)= \frac{e^2}{N}\int \frac{d^3 k}{(2 \pi)^3} \gamma^\mu S_F(p-k)\gamma^\nu \Delta_{\mu\nu}(k). \label{fermionA1}
\end{equation}
The first step is to use Eq.~\eqref{fermionbare} and Eq.~\eqref{fullfot2} in Eq.~(\ref{fermionA1}). On the other hand, the self-energy in the small-momentum limit, which is the revelant term in order to extract the form of the divergences, is written as \cite{fernandez, son}   
\begin{equation}
\begin{split}
\Sigma_{A_\mu}(p) &= \Sigma_{A_\mu}(p)\Bigm{|}_{p=0}+ \gamma^0 p_0 \frac{\partial \Sigma_{A_\mu}(p_0)}{\partial p_0}\Bigm{|}_{p_0=0} \\
&+ v_F \gamma^i p_i \frac{\partial \Sigma_{A_\mu}(p_i)}{\partial p_i}\Bigm{|}_{p_i=0 }+\cdots.
\end{split}
\end{equation}
After some calculations, (see  App. A of ref. \cite{fernandez}) it is possible to show that the  fermion self-energy, in the small-mass limit, is
\begin{equation}
\begin{split}
\Sigma_{A_\mu}(p) &= -\frac{2 \lambda}{\pi^2 N}\left[\gamma^0 p_0 f_1(\lambda) - v_F \gamma^i p_i f_2(\lambda) \right. \\ & \left. + m f_0(\lambda)\right]\ln\left(\frac{\Lambda}{\Lambda_0}\right) + \textrm{FT},
\end{split}
\label{selfenergyfoton}
\end{equation}
where $\textrm{FT}$ stands for finite terms, $\lambda =  e^2/ (16 v_F)=\pi \alpha/4$,
where $\alpha$ is the fine-structure constant,
\begin{equation}
f_0(\lambda) = \frac{2\cos^{-1}(\lambda)}{\sqrt{1-\lambda}},
\end{equation}

\begin{equation}
f_1(\lambda)= -\frac{2}{\lambda^2}\left[\pi -2 \lambda + \frac{(\lambda^2-2)}{\sqrt{1-\lambda^2}}\cos^{-1}(\lambda)\right],
\end{equation}
and
\begin{equation}
f_2(\lambda)= \frac{1}{\lambda^2}\left[\pi -2\lambda - 2\sqrt{1-\lambda^2}\cos^{-1}(\lambda)\right].
\end{equation}

\subsubsection{Self-energy due to the auxiliary fields}
The fermion self-energy due to the auxiliary fields is shown in Fig. \ref{selfenergy1}\textcolor{red}{.b} and its analytical expression is given by
\begin{equation}
\Sigma_{\{\varphi_l\}}(p)= \int \frac{d^3 k}{(2 \pi)^3} \Gamma^l S_F(p-k) \Gamma^{l'} \Delta_{l l'}^{\{\varphi_l\}}(k). \label{fermionaux}
\end{equation}
Here we use Eq.~\eqref{fermionbare} and Eq.~(\ref{fullauxiliary}) in Eq.~(\ref{fermionaux}). Thereafter, we make  the same reassignement of the momentum variables as before, i.e., we redefine the  external momentum as $p_\mu \rightarrow \bar{p}_{\mu}$, where $\bar{p}_{\mu}=(p_0,\bar{p}_i)$, and change the loop-integral variable as  $v_F k_i \rightarrow k_i$. Therefore, the self-energy is written as
\begin{equation}
\Sigma_{\{\varphi_l\}}(\bar{p})\!=\! \!\frac{1}{v_F^2}\!\! \int\!\!\! \frac{d^3 k}{(2 \pi)^3} \Gamma^l \frac{\gamma^\alpha (\bar{p}\!-\!k)_\alpha\!\!+\!m}{(\bar{p}-k)^2 +m^2} \Gamma^{l'} \Delta_{l l'}^{\{\varphi_l\}}(k).
\label{selfenergy}
\end{equation}
Similarly to the previous case, we expand the self-energy as
\begin{equation}
\Sigma_{\{\varphi_l\}}(\bar{p})\!=\!\!\Sigma_{\{\varphi_l\}}\!(\bar{p})\Bigm{|}_{\bar{p}=0}\!\!\! + \bar{p}_\mu \frac{\partial  \Sigma_{\{\varphi_l\}}(\bar{p})}{\partial \bar{p}_\mu} \Bigm{|}_{\bar{p}=0} + \cdots. \label{SEaux}
\end{equation}

Note that the lowest-order term in Eq.~(\ref{SEaux}) is
\begin{equation}
\Sigma_{\{\varphi_l\}}(\bar{p})\Bigm{|}_{\bar{p}=0} = \frac{m}{v_F^2} \int \frac{d^3 k}{(2 \pi)^3} \frac{\Gamma^l \Gamma^{l'}}{k^2 + m^2} \Delta_{l l'}^{\{\varphi_l\}}(k)
\label{Ozero}
\end{equation}
and the first-order term in $\bar{p}_\mu$ reads
\begin{equation}
\begin{split}
\bar{p}_\mu \frac{\partial \Sigma_{\{\varphi_l\}}(\bar{p}) }{\partial \bar{p}_\mu}\Bigm{|}_{\bar{p}=0} &= \frac{1}{v_F^2} \int \frac{d^3 k}{(2 \pi)^3} \Gamma^l \left\lbrace \frac{\gamma^\mu \bar{p}_\mu}{k^2 + m^2} \right. \\ & \left. - 2 \frac{\gamma^\alpha k_\alpha k^\mu \bar{p}_\mu}{\left[k^2 +m^2\right]^2} \right\rbrace \Gamma^{l'} \Delta_{l l'}^{\{\varphi_l\}}(k).
\label{Ofirst}
\end{split}
\end{equation}

For analyzing  the divergent parts of these expressions we may neglect  the $1/G_{\varphi_{l}}$ terms in the propagators of the auxiliary fields as they only give finite contributions.
We assume that $1/G_{\varphi_l} \ll \sqrt{\bar{p}^2}$ as an approximation in the auxiliary-field propagators for calculating the fermion self-energy. Let us take the  Thirring interaction as a concrete example, hence, $\{\varphi_l\}\rightarrow \varphi_2=\varphi_{\gamma_{\mu}}$. In this case, the zero-order term reads
\begin{equation}
\begin{split}
\Sigma_{\varphi_\mu}(\bar{p})\Bigm{|}_{\bar{p}=0}&=\frac{8}{N} m \int \frac{d^3 k}{(2 \pi)^3} \left\lbrace \frac{\gamma^\mu \gamma_\mu}{\left(k^2+m^2\right) }\right. \\ & \left. - \frac{\gamma^\mu  \gamma^\nu k_\mu  k_\nu}{(k^2 + m^2) k^2}\right\rbrace \frac{1}{\sqrt{k^2}}. \label{phimu0}
\end{split}
\end{equation}
Next, we use $\gamma^\mu \gamma_\mu = -3$	 and, given the Lorentz invariance on the integral, we change $k_\mu k_\nu \rightarrow g_{\mu \nu} k^2/3$. Using these conditions, Eq.~(\ref{phimu0}) yields
\begin{equation}
\Sigma_{\varphi_\mu}(\bar{p})\Bigm{|}_{\bar{p}=0}=-\frac{16}{N} m \int \frac{d^3 k}{(2 \pi)^3} \frac{1}{k^2 +m^2}\frac{1}{\sqrt{k^2}}.
\end{equation}
After applying the Feynman parametrization and cutt-off regularization, within the small-mass limit, we find the zero-order term, namely,
\begin{equation}
\Sigma_{\varphi_\mu}(\bar{p}) \Bigm{|}_{\bar{p} = 0} =-\frac{8}{\pi^2 N} m \ln\left(\frac{\Lambda}{\Lambda_0}\right).
\label{resultadoordenzerothirring}
\end{equation}

Next, let us calculate the first-order term. From the  expansion given by  Eq. \eqref{Ofirst}, we find
\begin{equation}
\begin{split}
\bar{p}_\beta \frac{\Sigma_{\varphi_\mu}(\bar{p})}{\partial \bar{p}_\beta} \Bigm{|}_{\bar{p}=0} &= \frac{8}{N}\int \frac{d^3 k}{(2 \pi)^3} \gamma^\mu\left\lbrace \frac{\gamma^\beta \bar{p}_\beta}{k^2+m^2} \right. \\ & \left. -2\frac{\gamma^\alpha k_\alpha k^\beta \bar{p}_\beta}{\left(k^2+m^2\right)^2}\right\rbrace\gamma^\nu\frac{\mathbb{K_{\mu \nu}}}{\sqrt{k^2}}.
\end{split}
\label{ordem1thirring}
\end{equation}
We shall follow the same steps as before. Here, however, for the second integral in rhs of Eq.~(\ref{ordem1thirring}), we use $k_\alpha k^\beta k_\mu k_\nu \rightarrow \left(\delta_\alpha^\beta \delta _{\mu \nu} + \delta_{\alpha\mu} \delta _{\nu}^\beta+\delta_{\alpha \nu} \delta_{\mu}^{\beta} \right) k^4/15$, because of the Lorentz invariance in the loop integral. Therefore, we obtain 
\begin{equation}
\bar{p}_\beta \frac{\partial \Sigma_{\varphi_\mu}(\bar{p})}{\partial \bar{p}_\beta}\Bigm{|}_{\bar{p}=0}= - \frac{8}{3 \pi^2 N} \gamma^\beta \bar{p}_\beta  \ln \left(\frac{\Lambda}{\Lambda_0}\right) + \textrm{FT}, 
\label{resultadorden1thirring}
\end{equation}
being $\textrm{FT}$ the finite terms. From Eq.~(\ref{resultadoordenzerothirring}) and Eq.~(\ref{resultadorden1thirring}), we find the whole contribution of the Thirring interaction to the the fermion self-energy, given by  
\begin{equation}
\Sigma_{\varphi_{\mu}}(\bar{p})= - \frac{8}{3 \pi^2 N} \left\lbrace \gamma^\mu \bar{p}_\mu + 3 m \right\rbrace \ln\left(\frac{\Lambda}{\Lambda_0}\right) + \textrm{FT}.
\label{thirringselfenergy}
\end{equation}

After doing the same procedure for the other interactions, we find
\begin{equation}
\begin{split}
\Sigma_{\varphi_\mathbbm{1}}(\bar{p}) & = \! \frac{2}{3 \pi^2 N}\! \left\lbrace \! \gamma^\mu \bar{p}_\mu\!+\! 3 m \right\rbrace \ln\!\left(\frac{\Lambda}{\Lambda_0}\right)\!+\! \textrm{FT},\\\Sigma_{\varphi_{3(5)}}(\bar{p}) & \!=\!  \frac{2}{3 \pi^2 N} \!\left\lbrace\! \gamma^\mu \bar{p}_\mu \!- \!3 m \right\rbrace \ln\left(\frac{\Lambda}{\Lambda_0}\right)\!+\! \textrm{FT},\\
\Sigma_{\varphi_{\mu3(5)}}\!(\bar{p}) & \!=\! -\frac{8}{3 \pi^2 N} \! \left\lbrace \! \gamma^\mu \bar{p}_\mu \!-\!3 m \right\rbrace \!\ln\!\left(\frac{\Lambda}{\Lambda_0}\right)\!+\! \textrm{FT},\\
\Sigma_{\varphi_{35}}(\bar{p}) & \! =\!  -\frac{2}{3 \pi^2 N} \!\left\lbrace \! \gamma^\mu \bar{p}_\mu \!+\! 3 m \right\rbrace \! \ln\!\left(\frac{\Lambda}{\Lambda_0}\right)\!+\! \textrm{FT},
\label{auto4f}
\end{split}
\end{equation}
and
\begin{equation}
\Sigma_{\varphi_{\mu 3 5}}\!(\bar{p}) \!=\!  -\frac{8}{3 \pi^2 N} \! \left\lbrace \gamma^\mu \bar{p}_\mu \!+\! 3  m \right\rbrace \ln\left(\frac{\Lambda}{\Lambda_0}\right)\!+\! \textrm{FT}.
\label{auto4f2}
\end{equation}

\section{Renormalization group}

In general grounds, the renormalization group equation has so many anomalous dimensions as fields in the lagrangian. However, because that the vacuum polarization tensors $\Pi^{\mu \nu}(p)$ and $\Pi^{ll'}_{\varphi_l}(p)$ are finite in the dimensional regularization scheme, we conclude that $\gamma_{A_\mu}$, $\beta_e$,  and  $\gamma_{\varphi_l}$ vanish. Furthermore, the beta functions for the coupling constants $\beta_G$ do not appear in our renormalization group equation due to the approximation $1/G_l \ll \sqrt{p^2}$ we have considered before. Having these assumption in mind, our renormalization group equation is written as 
\begin{equation}
\left[\Lambda \frac{\partial}{\partial \Lambda} \!+\!\beta_{v_F} \frac{\partial}{\partial v_F} \!+\! \beta_m \frac{\partial}{\partial m}\!-\! N_F \gamma_F\right]\!\! \Gamma^{(N_F, \dots)}\!= 0,
\label{egr} 
\end{equation}
where $\Gamma^{N_F, \dots} = \Gamma^{(N_F,N_A,N_{\varphi_l})}(p_1,\dots,p_N)$ are the renormalized vertex functions, ($N_F, N_A, N_{\varphi_l}$) are the number of external lines of fermion, gauge and auxiliary fields, respectively. The beta functions of $v_F$ and $m$ parameters are $\beta_{v_F}=\Lambda \frac{\partial v_F}{\partial \Lambda}$  and $\beta_m=\Lambda \frac{\partial m}{\partial \Lambda}$, respectively. The anomalous dimension of the fermion field is $\gamma_F = \Lambda \frac{\partial}{\partial \Lambda} \left(\ln Z_{\psi} \right)$, where $Z_{\psi}$ is the wave function renormalization.

The two-point function for the fermion field is 
\begin{equation}
\Gamma^{(2)}= \left(\gamma^0p_0 + v_F \gamma^i p_i - m\right) +\Sigma_{A_\mu}(p) +  \Sigma_{l}^{\{\varphi_l\}}(p),
\label{functionvertex}
\end{equation}
where the contribution of the gauge field $\Sigma_{A_\mu}(p)$ is
\begin{equation}
\Sigma_{A_\mu}(p) = \left[a_1\gamma^0p_0 + a_2  v_F \gamma^i p_i  + a_3 m \right]\ln\left(\frac{\Lambda}{\Lambda_0}\right),
\end{equation}
the coefficients $a_{1,2,3}$ are easily obtained from Eq. \eqref{selfenergyfoton}. 
After recovering $\bar{p} \rightarrow p = p_0 + v_Fp_i $, we find the contribution of the auxiliary fields, i.e, 
\begin{equation}
\!\!\!\Sigma_{l}^{\{\varphi_l\}}\!(p) \!\!=\!\! \left[b_1^{\{\varphi_l\}} \!\gamma^0 p_0 \!+\! b^{\{\varphi_l\}}_1\! v_F \gamma^i p_i \!+\!b^{\{\varphi_l\}}_2 m\right]\! \ln\!\left(\!\frac{\Lambda}{\Lambda_0}\!\right)\!,
\end{equation}
where the coefficients $b_{1,2}$ are obtained from Eq.~\eqref{thirringselfenergy},  \eqref{auto4f}, and \eqref{auto4f2}. In the large-$N$ expansion, we may write the beta functions as $\beta_a = \beta_a^{(1)} + \frac{1}{N} \beta_a^{(2)} + \cdots$, with $a=v_F,\, m$, and the anomalous dimension as $\gamma_F = \gamma_F^{(1)} + \frac{1}{N}\gamma_F^{(2)}+ \cdots$. Thereafter, we replace Eq.~\eqref{functionvertex} in Eq.~\eqref{egr} and, after some algebra, we obtain
\begin{equation}
\gamma_F = \frac{1}{2}\left(a_1 + \sum_l b_1^{\{\varphi_l\}}\right),
\end{equation}
\begin{equation}
\beta_{v_F}= v_F\left(a_1-a_2\right),
\end{equation}
and
\begin{equation}
\beta_{m}= m\left[a_1+a_3+ \sum_l\left(b_1^{\{\varphi_l\}}+b_2^{\{\varphi_l\}}\right)\right].
\label{massgeral}
\end{equation}
Using the coefficients $a_{1,2,3}$ and $b_{1,2}$, we obtain
\begin{equation}
\begin{split}
\gamma_F & = -\frac{2}{\pi^2 N } \left[2+\frac{2-\lambda^2}{\lambda \sqrt{1-\lambda^2}}\cos^{-1}\left(\lambda\right)-\frac{\pi}{\lambda}\right] \\ & - \frac{28}{3 \pi^2 N},
\label{anomalous}
\end{split}
\end{equation}
\begin{equation}
\beta_{v_F}=-\frac{4}{\pi^2 N}v_F \left[1+\frac{\cos^{-1 }(\lambda)}{\lambda \sqrt{1-\lambda^2}}-\frac{\pi}{2 \lambda}\right],
\end{equation}
and
\begin{equation}
\begin{split}
\beta_m &=\! -\frac{2 m}{\pi^2 N} \left[4\!+\!\frac{4 \cos^{-1}\left(\lambda\right)}{\lambda\sqrt{1-\lambda^2}}-\frac{2\pi}{\lambda}\right]+ \!\! \sum_{\{b_{4F}\}}m \,b_{4F},
\label{mass}
\end{split}
\end{equation} 
where $b_{4F}\equiv b_1^{\{\varphi_l\}}+b_2^{\{\varphi_l\}}$ is the contribution, due to the four-fermion interactions, for the beta function of the mass. These are, in principle, different for each $\bar\psi\Gamma_l \psi$-term. Notice however that they do not depend on the couplings $1/G_{l}$. In fact, by considering the high momenta expansions for the auxiliary field propagators we may verify  that  terms containing these parameters are actually finite. In table I, we summarize all of the possible values of $b_{4F}$ generated by each individual interaction. 

\begin{table}[H]
\centering
\begin{tabular}{|c|c|}
\hline
The Four-Fermion Interactions & The contribution $b_{4F}$  \\ 
 \hline
 $(\bar{\psi}\psi)^2$ & $8/(3 \pi^2 N)$  \\  
 \hline
 $(\bar{\psi}\gamma^\mu\psi)^2$ & $- 32/(3\pi^2 N)$  \\
 \hline
  $(\bar{\psi}\gamma^{3}\psi)^2$ & $- 4/(3 \pi^2 N)$  \\
 \hline
  $(\bar{\psi}\gamma^{5}\psi)^2$ & $- 4/(3 \pi^2 N)$  \\
 \hline
   $(\bar{\psi}\gamma^{3}\gamma^5\psi)^2$ & $- 8/(3 \pi^2 N)$  \\
 \hline
   $(\bar{\psi}\gamma^{\mu}\gamma^{3}\psi)^2$ & $16/(3 \pi^2 N)$  \\
 \hline
  $(\bar{\psi}\gamma^{\mu}\gamma^{5}\psi)^2$ & $16/(3 \pi^2 N)$  \\
 \hline
   $(\bar{\psi}\gamma^{\mu}\gamma^3 \gamma^5\psi)^2$ & $- 32/(3 \pi^2 N)$  \\
 \hline
\end{tabular}
\caption{The $b_{4F}$-term of each four-fermion interaction. This table gives the contribution of each four-fermion interaction to the beta function of the mass, given in Eq. \eqref{massgeral} by $b_1^{\{\varphi_l\}}+b_1^{\{\varphi_l\}}$. These contributions are calculated in the small-mass limit, where $m^2 \ll p^2$ and for $1/G_l \ll p$.}
\end{table}

\subsection{Mass Renormalization}
\begin{figure}[H]
\centering
\includegraphics[scale=0.65]{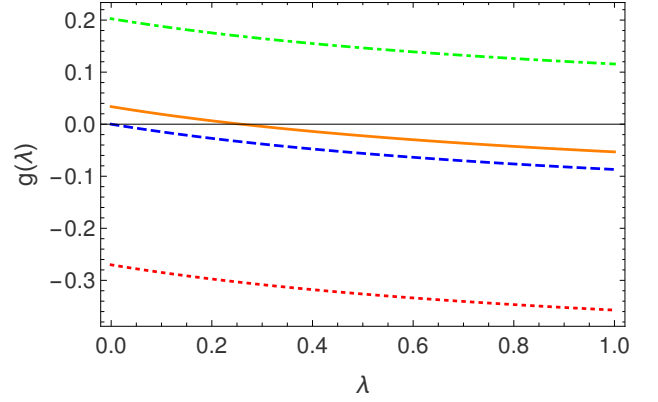}
\caption{The function $g(\lambda)$ in the interval $\lambda \in \left[0.1,1\right]$ with $N=4$. We plot Eq.~(\ref{glambda}) for four different combinations of the four-fermion interactions. The continuos line is obtained from the $(\bar{\psi}\psi)^2$ and $(\bar{\psi}\gamma^{3}\psi)^2$ interactions, which provides $\sum_{\{b_{4F}\}}b_{4F}=1/3 \pi^2$. In this case, we find $\lambda^{\rm min}_c= 0.26$. The dashed line is the combination of $(\bar{\psi}\psi)^2$ and $(\bar{\psi}\gamma^3 \gamma^5\psi)^2$  interactions, where $\sum_{\{b_{4F}\}}b_{4f}=0$. The dotted line is obtained from the $(\bar{\psi}\gamma^\mu\psi)^2$ interaction, where $b_{4F}=-8/3\pi^2$. The dashed-dotted line is obtained from the $(\bar{\psi}\psi)^2$ and $(\bar{\psi}\gamma^\mu \gamma^3\psi)^2$ interactions, where $\sum_{\{b_{4F}\}}b_{4F}=2/\pi^2$. Note that only for the continuos line we have a critical point.} \label{figglambda}
\end{figure}

\begin{figure}[H]
\centering
\includegraphics[scale=0.65]{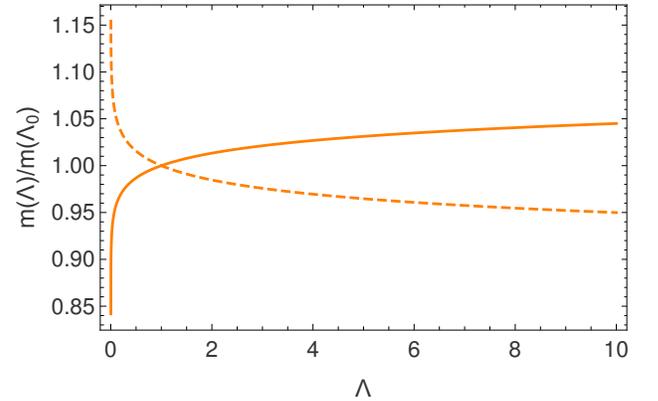}
\caption{The behavior of $m(\Lambda)$. We plot Eq.~(\ref{mLambda}) considering the $(\bar{\psi}\psi)^2$ and $(\bar{\psi}\gamma^3\psi)^2$ interactions, which provides $\lambda^{\rm min}_c=0.26$. For the line, we use $\lambda = 0.1$ while for the dashed line we use $\lambda=0.5$. We consider $N=4$ for both curves.} \label{fig4}
\end{figure}

We obtain the renormalized mass through the beta function as 
\begin{equation}
\Lambda \frac{\partial  m}{\partial \Lambda} = \beta_m \label{betamx}
\end{equation}
with Eq.~\eqref{mass}, where the renormalized mass depends on the energy scale $\Lambda$. After solving Eq.~(\ref{betamx}) for $m(\Lambda)$, it follows that
\begin{equation}
m\left(\Lambda\right) = m\left(\Lambda_0\right) \left(\frac{\Lambda}{\Lambda_0}\right)^{g\left(\lambda\right)}, \label{mLambda}
\end{equation}
where 
\begin{equation}
g(\lambda)=-\frac{2 }{\pi^2 N} \left[4+\frac{4 \cos^{-1}\left(\lambda\right)}{\lambda\sqrt{1-\lambda^2}}\right]+ \sum_{\{b_{4F}\}} b_{4F}. \label{glambda}
\end{equation}
From Eq.~(\ref{mLambda}) and Eq.~(\ref{glambda}), we conclude that the contribution of the $b_{4F}$-terms, generated by the four-fermion interactions, modifies the behavior of the renormalized mass, because they may change the sign of the function $g(\lambda)$, as shown in Fig.~\ref{figglambda}. In Fig.~\ref{fig4}, we plot the function $m(\Lambda)$. In general, there are three possible cases, namely: (A) $g(\lambda)<0$ for any $\lambda$, (B) $g(\lambda)>0$ for any $\lambda$, and (C) where either $g(\lambda)>0$ for $\lambda>\lambda_c$ or $g(\lambda)<0$ for $\lambda<\lambda_c$. The critical point $\lambda_c$ is obtained from $g(\lambda_c)=0$. Obviously, only the case (C) allow us to control the renormalized mass by tunning the value of $\lambda$.

\begin{figure}[H]
\centering
\includegraphics[scale=0.65]{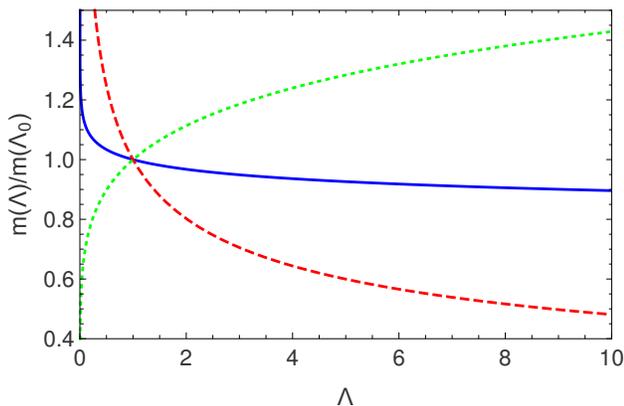}
\caption{ The behavior of $m(\Lambda)$. We plot Eq.~(\ref{mLambda}) with $N=4$ and $\lambda=0.4$. The continuos line is obtained from the $(\bar{\psi}\psi)^2$ and $(\bar{\psi}\gamma^3\gamma^5\psi)^2$ interactions (the same curve holds when we replace $\gamma_3$ by $\gamma_5$). The dashed line is obtained from the Thirring interaction $(\bar{\psi}\gamma^\mu\psi)^2$. The dotted line is obtained from $(\bar{\psi}\psi)^2$ and $(\bar{\psi}\gamma^\mu \gamma^3 \psi)^2$ interactions.} \label{fig5}
\end{figure}

Next, let us consider the case (A). This is the regime where the renormalized mass is fully controlled by the electromagnetic interactions. Therefore, the sum over the $b_{4F}$-term vanishes. In Fig. \ref{fig5}, we show a plot for such possibility. From table I, we conclude that there are six different combinations that fulfil this criteria. These are: (A.1) $\left(\bar{\psi}\psi\right)^2+\left(\bar{\psi}\gamma^3 \gamma^5\psi\right)^2$, (A.2) $\left(\bar{\psi}\psi\right)^2+\left(\bar{\psi}\gamma^3\psi\right)^2+\left(\bar{\psi}\gamma^5\psi\right)^2$, (A.3) $\left(\bar{\psi}\gamma^\mu\psi\right)^2+\left(\bar{\psi}\gamma^\mu\gamma^3 \psi\right)^2+\left(\bar{\psi}\gamma^\mu\gamma^5 \psi\right)^2$ (where $\left(\bar{\psi}\gamma^\mu \psi\right)^2$ can be replaced by $\left(\bar{\psi}\gamma^\mu \gamma^3 \gamma^5 \psi\right)^2$), and (A.4) $\left(\bar{\psi}\gamma^\mu \gamma^3\psi\right)^2+\left(\bar{\psi}\gamma^3 \gamma^5\psi\right)^2+\left(\bar{\psi}\gamma^3\psi\right)^2+\left(\bar{\psi}\gamma^5\psi\right)^2$ (where $\left(\bar{\psi}\gamma^\mu \gamma^3 \psi\right)^2$ can be replaced by $\left(\bar{\psi}\gamma^\mu \gamma^5 \psi\right)^2$). In the case (A), we conclude that $m(\Lambda)\rightarrow 0$ as $\Lambda\rightarrow \infty$.

In the case (B), we need combinations that always provide $g(\lambda)$ positive. In this regime, the influence of the four-fermion interactions is dominant over the contribution of the electromagnetic interactions. Therefore, the sum over the $b_{4F}$-term must be larger than the first term in the rhs of Eq.~(\ref{glambda}), for any $\lambda$. In Fig. \ref{fig5}, we show a plot for such possibility. From table I, we conclude that there are seven different combinations that fulfil this criteria. These are: (B.1) $\left(\bar{\psi}\gamma^\mu\gamma^3\psi\right)^2+\left(\bar{\psi}\gamma^\mu\gamma^5\psi\right)^2$, (B.2) $\left(\bar{\psi}\psi\right)^2+\left(\bar{\psi}\gamma^\mu\gamma^3\psi\right)^2$ (where $\left(\bar{\psi}\gamma^\mu\gamma^3\psi\right)^2$ can be replaced by $\left(\bar{\psi}\gamma^\mu\gamma^5\psi\right)^2$), (B.3) $\left(\bar{\psi}\gamma^\mu\gamma^3\psi\right)^2+\left(\bar{\psi}\gamma^3\psi\right)^2$, and (B.4) $\left(\bar{\psi}\gamma^\mu\gamma^5\psi\right)^2+\left(\bar{\psi}\gamma^3\psi\right)^2$ (in the last two combinations $\left(\bar{\psi}\gamma^3\psi\right)^2$ can replaced by $\left(\bar{\psi}\gamma^5\psi\right)^2$). In the case (B), we conclude that $m(\Lambda)\rightarrow \infty$ as $\Lambda\rightarrow \infty$.

In the case (C), the sign of $g(\lambda)$ changes after crossing the point $\lambda_c$. In this regime, the renormalized mass is described by the competition of electromagnetic and four-fermion interactions, where both of them are relevant. We find two possible values for the critical coupling constant, namely, $\lambda^{{\rm max}}_c=0.66$ and $\lambda^{{\rm min}}_c=0.26$. From table I, we find seven combinations that provide $\lambda^{{\rm max}}_c$, given by: (C.1.A) $(\bar\psi\psi)^2$, (C.2.A) $\left(\bar{\psi}\gamma^\mu \gamma^3 \psi\right)^2+\left(\bar{\psi}\gamma^3 \gamma^5\psi\right)^2$, (C.3.A) $\left(\bar{\psi}\psi\right)^2 + \left(\bar{\psi}\gamma^\mu \gamma^3\psi\right)^2+\left(\bar{\psi}\gamma^3 \gamma^5\psi\right)^2+\left(\bar{\psi}\gamma^3 \psi\right)^2+\left(\bar{\psi}\gamma^5 \psi\right)^2$, and (C.4.A) $\left(\bar{\psi}\gamma^\mu \gamma^3\psi\right)^2+\left(\bar{\psi}\gamma^3 \psi\right)^2+\left(\bar{\psi}\gamma^5 \psi\right)^2$ (in each of the previous combinations we can change $\left(\bar{\psi}\gamma^\mu \gamma^3 \psi\right)^2$ by $\left(\bar{\psi}\gamma^\mu \gamma^5 \psi\right)^2$). On the other hand, for finding $\lambda^{{\rm min}}_c$, there are six possibilities, namely, (C.1.B) $(\bar\psi \psi)^2+(\bar{\psi}\gamma^{3}\psi)^2$ and (C.2.B) $(\bar\psi \psi)^2+(\bar{\psi}\gamma^{5}\psi)^2$, (C.3.B) $\left(\bar{\psi}\gamma^\mu \gamma^3\psi\right)^2 +\left(\bar{\psi}\gamma^3 \gamma^5\psi\right)^2+\left(\bar{\psi}\gamma^3\psi\right)^2$, (C.4.B) $\left(\bar{\psi}\gamma^\mu \gamma^3\psi\right)^2 +\left(\bar{\psi}\gamma^3 \gamma^5\psi\right)^2+\left(\bar{\psi}\gamma^5\psi\right)^2$, (C.5.B) $\left(\bar{\psi}\gamma^\mu \gamma^5\psi\right)^2 +\left(\bar{\psi}\gamma^3 \gamma^5\psi\right)^2+\left(\bar{\psi}\gamma^3\psi\right)^2$, and (C.6.B) $\left(\bar{\psi}\gamma^\mu \gamma^5\psi\right)^2 +\left(\bar{\psi}\gamma^3 \gamma^5\psi\right)^2+\left(\bar{\psi}\gamma^5\psi\right)^2$. The case (C) clearly provides two possible asymptotic behaviors for $m(\Lambda)$, see Fig.~\ref{fig5}.

In Ref.~\cite{fernandez}, it has been shown that the combination of electromagnetic and Gross-Neveu interactions yields $\lambda^{\rm max}_c=0.66$, which is our case (C.1.A). We believe that combinations with a \textit{minimal critical coupling constant} $\lambda^{\rm min}_c=0.26$, see Fig.~\ref{fig4}, are likely to provide an easier controlling of the renormalized mass. Indeed,  because screening effects, due to the substrates, decrease the value of $\lambda$, hence, the phase when $\lambda>\lambda_c$ becomes harder to achieve experimentally. From the experimental point of view, it is possible to relate the energy scale $\Lambda$ with the electronic density $n$ (the number of electrons by unit of surface area) by using the scaling law $\Lambda \rightarrow n^{1/2}$ \cite{sarman}. The value of $n$ is controlled by a gate voltage \cite{fernandez}.  We believe that our results may be relevant for describing a more realistic process of mass renormalization. Obviously, the four-fermion interactions should be related with microscopic interactions, such as mechanical vibrations, impurities, and disorder in the honeycomb lattice.

\section{SUMMARY AND OUTLOOK}

The experimental realization of two-dimensional materials, where the quasiparticles obey a Dirac-like equation, allow us to consider a quantum-electrodynamical approach in order to describe electronic interactions in these systems. In particular, the experimental observation \cite{geim} of the Fermi velocity renormalization \cite{vozmediano} in graphene confirms that electronic interactions are indeed relevant. Recently, the description of the band gap renormalization \cite{fernandez} in WSe$_2$ \cite{WSe} and MoS$_2$ \cite{MoS} increases this window of possible applications, using standard renormalization group equations, as in Ref.~\cite{foster}. Within a nonperturbative regime, one can also consider the description of excitonic spectrum \cite{Exc}, dynamical mass generation \cite{popovici}, and the realization of parity anomaly \cite{PRX} through a quantum valley Hall effect. Beyond these regimes, one can consider the microscopic interactions by taking models that simultaneously describe both electromagnetic and four-fermion interactions. These cases, however, have been less discussed in literature \cite{EbertFF}.

In this work we  gave a step forward in this picture by considering an effective low-energy model that is suitable for calculating the effects of both electromagnetic and  the generalized four-fermion interactions with $O(4)$ symmetry. As a concrete application, we calculated the renormalized mass $m(\Lambda)$, within the large-$N$ approximation. This may be measured by looking at the energy gap between the valence (negative energy) and conduction (positive energy) bands at the valleys of the honeycomb lattice \cite{WSe,MoS}. For the sake of comparison with the experimental data, we may replace the energy scale $\Lambda$ by the electron density $n$, through the transform $\Lambda \rightarrow n^{1/2}$, which is true for two-dimensional electrons \cite{vozmediano,fernandez}. Our result shows that an ultraviolet fixed point $\lambda_c$ is generated, implying that $m(\Lambda)$ does not renormalizes at $\lambda=\lambda_c$. Thereafter, we find that there exist two possible values for $\lambda_c$, namely, the maximal value $\lambda^{\rm max}_c=0.66$ and the minimal value $\lambda^{\rm min}_c=0.26$ (this does not depends on the constant $N$). The kind of value we find depends on the combinations of four-fermion interactions we are considering in the initial model. This provides a possible tuning mechanism for the renormalized mass, because the behavior of $m(\Lambda)$ changes when $\lambda=\pi \alpha/4$ is either larger or less than $\lambda_c$.

The model presented here is also suitable for investigating the ultrarelativistic limit of the Dirac-like materials, where $v_F(n)\rightarrow c$ as $n\rightarrow 0$, where $c$ is the light velocity. Because our current results only describe the regime where $v_F(n)\ll c$ (the static limit), it would be interesting to understand the behavior of the renormalized mass in the dynamical limit. We shall consider this generalization elsewhere.

\section*{Acknowledgement}
L. F. is partially supported by Coordenação de Aperfeiçoamento de Pessoal de Nível Superior Brasil (CAPES), finance code 001. V. S. A. and L. O. N. are partially supported by Conselho Nacional de Desenvolvimento Científico e Tecnológico (CNPq) and by CAPES/NUFFIC, finance code 0112. F. P. acknowledge the financial support from Dirección De Investigación De La Universidad De La Frontera Grant No. DI20-0005.

\appendix
\numberwithin{equation}{section}
\textbf{\section{\textbf{Vacuum polarization tensor of four-fermion interactions}}\label{A}}

Equation \eqref{Polarization4f} represents the general form of the one-loop quantum correction to the auxiliary-field propagators. Here, we provide a few details of the computation of this term for the case of the Thirring interaction where the vertex is $\varphi_\mu \bar{\psi}\gamma^{\mu}\psi$. In this case, we have
\begin{equation}
\Pi^{\mu \nu}_{\varphi_\mu}(\bar{p})\! =\!\!- \frac{1}{N v_F^2} {\rm Tr}\!\! \int\!\! \frac{d^3 k}{(2 \pi)^3} \gamma^\mu S_F\!\left(\bar{p}+k\right) \gamma^\nu S_F \left( k \right).  \label{A1}
\end{equation}
Next, we use Eq.~(\ref{fermionbare}) and the following trace operations over the Dirac matrices, namely, 
\begin{align}
{\rm Tr}[\gamma^\mu \gamma^\nu]  &= - 4 \delta^{\mu \nu}, \\
{\rm Tr}\left[\gamma^\mu \gamma^\alpha \gamma^\nu \gamma^\beta\right] &= 4\! \left(\delta^{\mu \alpha} \!\delta^{\nu \beta}\!\!-\delta^{\mu \nu} \delta^{\alpha \beta} + \delta^{\mu \beta} \delta^{\nu \alpha}\right),
\end{align}
which are useful properties to expand the numerator of Eq.~(\ref{A1}). After calculating the trace over the Dirac matrices, this numerator reads 
\begin{equation}
(\bar{p}+k)^\mu k^\nu+(\bar{p}+k)^\nu k^\mu - \delta^{\mu \nu}\left[(\bar{p}+k)\cdot k +m^2\right].
\end{equation}
On the other hand, we use the Feynman parametrization in the denominator, which becomes equal to
\begin{equation}
\left[(k+x \bar{p})^2 +x(1+x)\bar{p}^2 +m^2\right]^2.
\end{equation}
Thereafter, in order to eliminate symmetric-loop integrals, we made a variable change $k \rightarrow k - x\bar{p}$ and, by  using  Lorentz invariance, we finally find a simplified equation for $\Pi^{\mu \nu}_{\varphi_\mu}(\bar{p})$, namely,
\begin{equation}
\begin{split}
\Pi^{\mu \nu}_{\varphi_\mu}(\bar{p})&= - \frac{4}{v_F^2} \int^1_0 dx \left\lbrace \int \frac{d^3 k}{(2 \pi)^3} \frac{ \left(\frac{2}{3}-1 \right) \delta^{\mu \nu}k^2}{\left[k^2 + \Delta_1 \right]^2} \right. 
\\ 
& \left.\!\!\! \int \!\!\frac{d^3 k}{(2 \pi)^3} \frac{\left(\!\delta^{\mu \nu}\!\!-\!\! 2 \frac{\bar{p}^\mu \bar{p}^\nu}{\bar{p}^2}\right)\! x(1\!-\!x)\bar{p}^2\!\!-\!\delta^{\mu \nu}m^2}{\left[k^2 +\Delta_1\right]^2}\right\rbrace
\end{split}
\end{equation}  
with $\Delta_1=x(1-x)\bar{p}^2 +m^2$. After solving the integrals, using the dimensional regularization scheme, we find
\begin{equation}
\begin{split}
\Pi^{\mu \nu}_{\varphi_\mu}(\bar{p})&= -\frac{\bar{p}^2}{\pi v_F^2} \left\lbrace \frac{\sqrt{m^2}}{2 \bar{p}^2} + \frac{\bar{p}^2 -4 m^2}{4 \bar{p}^2 \sqrt{\bar{p}^2}} \right. 
\\
& \left. \times \arcsin\left[\sqrt{\frac{\bar{p}^2}{\bar{p}^2+4 m^2}}\right]\right\rbrace \mathbb{\bar{P}}^{\mu \nu}.
\end{split}
\end{equation}

 In the cases of other auxiliary fields, we use that $\gamma^{3}$ and $\gamma^5$ anti-commute with $\gamma^\mu$ and between them, furthermore, $(\gamma^{3})^2=(\gamma^5)^2 = \mathbbm{1}$. Hence, it follows some useful properties, given by
\begin{align}
&{\rm Tr}[\gamma^{3 (5)} \gamma^{3 (5)}]  =  4, \label{33} \\
& {\rm Tr}\left[\gamma^{3 (5)} \gamma^\alpha \gamma^{3 (5)} \gamma^\beta \right] = 4 \delta^{\alpha \beta}, \label{3alpha}\\
&{\rm Tr}\left[\gamma^3 \gamma^5 \gamma^3 \gamma^5 \right] = -4, \label{35} \\
&{\rm Tr} \left[ \gamma^\mu \gamma^{3(5)} \gamma^\alpha \gamma^\nu \gamma^{3(5)} \gamma^\beta \right] ={\rm Tr}\left[\gamma^\mu \gamma^\alpha \gamma^\nu \gamma^\beta\right], \label{mu3} \\
&{\rm Tr} \left[\gamma^3 \gamma^5 \gamma^\alpha \gamma^3 \gamma^5 \gamma^\beta \right] = 4 \delta^{\alpha \beta} \label{35alpha} \\
&{\rm Tr}\left[ \gamma^\mu \gamma^3 \gamma^5 \gamma^\alpha \gamma^\nu \gamma^3 \gamma^5 \gamma^\beta \right] = - {\rm Tr}\left[\gamma^\mu \gamma^\alpha \gamma^\nu \gamma^\beta\right]. \label{mu35}
\end{align}

We obtain $\Pi_{\varphi_3}(\bar{p})$ and $\Pi_{\varphi_5}(\bar{p})$ using $\Gamma^l=\gamma^3$ or $\gamma^5$ respectively, in Eq. \eqref{Polarization4f}. Then, we implement the same procedure for solve Eq. \eqref{A1} together with Eqs. \eqref{33} and \eqref{3alpha}, of form that 
\begin{equation}
\Pi_{\varphi_3}\!(\bar{p})\!=\!\!\frac{\sqrt{\bar{p}^2}}{\pi v_F^2}\!\left[\!\sqrt{\frac{m^2}{\bar{p}^2}}\!+\!\frac{1}{2}\!\arcsin\left(\!\!\sqrt{\frac{\bar{p}^2}{\bar{p}^2+4m^2}}\right)\!\right],
\end{equation}
for $\Pi_{\varphi_5}(\bar{p})$ we have the same previous result. We use $\Gamma^l=\gamma^{\mu }\gamma^3$ (or $\gamma^{\mu }\gamma^5$) in Eq. \eqref{Polarization4f} and Eqs. \eqref{3alpha} and \eqref{mu3} for we obtain $\Pi_{\varphi_{\mu 3}}(\bar{p})$ (or $\Pi_{\varphi_{\mu 5}}(\bar{p})$), being
\begin{equation}
\begin{split}
\Pi_{\varphi_{\mu 3}}^{\mu \nu}&(\bar{p})  = - \frac{\sqrt{\bar{p}^2}}{2 \pi v_F^2}\left\lbrace \left[ \sqrt{\frac{m^2}{\bar{p}^2}}+\frac{\bar{p}^2+4 m^2}{2 \bar{p}^2} \right. \right. \\ &\left. \left. \times  \arcsin\left(\!\!\sqrt{\!\frac{\bar{p}^2}{\bar{p}^2+4m^2}}\,\right)\! \right] \mathbb{\bar{P}}^{\mu \nu} \!+2 \frac{m^2}{\bar{p}^2}\delta^{\mu \nu}\right\rbrace,
\end{split}
\end{equation}
using $\Gamma^l = \gamma^3 \gamma^5$ in Eq. \eqref{Polarization4f} together with Eqs. \eqref{35} and \eqref{35alpha} we obtain
\begin{equation}
\begin{split}
\Pi_{35}(\bar{p})&=-\frac{\sqrt{\bar{p}^2}}{\pi v_F^2} \left[\sqrt{\frac{m^2}{\bar{p}^2}}+\frac{\bar{p}^2+4m^2}{2 \bar{p}^2} \right. \\ & \left. \times \arcsin\left(\sqrt{\frac{\bar{p}^2}{\bar{p}^2 +4m^2}}\right)\right]
\end{split},
\end{equation}
by last we may obtain $\Pi_{\varphi_{\mu 3 5}}^{\mu \nu}(\bar{p})$ by replacing  $\Gamma^l$ by $\gamma^\mu \gamma^3 \gamma^5$ in Eq. \eqref{Polarization4f} and using the trace operation given by Eqs. \eqref{35alpha} and \eqref{mu35}, so
\begin{equation}
\begin{split}
\Pi^{\mu \nu}_{\varphi_{\mu 3 5}}(\bar{p}) & = \frac{\bar{p}^2}{\pi v_F^2}\left[\frac{\sqrt{m^2}}{2 \bar{p}^2}+\frac{\bar{p}^2 - 4 m^2}{4 \bar{p}^2 \sqrt{\bar{p}^2}} \right. \\ & \left. \times \arcsin\left(\sqrt{\frac{\bar{p}^2}{\bar{p}^2+4m^2}}\right) \right] \mathbb{\bar{P}}^{\mu \nu}.
\end{split}
\end{equation}

\end{document}